\documentclass[twocolumn,showpacs,preprintnumbers,amsmath,amssymb,floatfix]{revtex4}

\usepackage{graphicx}
\usepackage{dcolumn}
\usepackage{bm}

\newcommand{\beq}{\begin{equation}}
\newcommand{\eeq}{\end{equation}}
\newcommand{\bey}{\begin{eqnarray}}
\newcommand{\eey}{\end{eqnarray}}

\newcommand{\grad}{{\bf \nabla}}

\begin{document}

\preprint{APS/123-QED}

\title{Comparing different realizations of modified Newtonian dynamics: \\ virial theorem and elliptical shells}

\author{HongSheng Zhao} \affiliation{SUPA, School of Physics and Astronomy, University of St Andrews, KY16 9SS, Fife, UK}%

\author{Benoit Famaey} 
\affiliation{Observatoire Astronomique,  Universit\'e de Strasbourg, CNRS UMR 7550, F-67000 Strasbourg, France}

\date{\today}

\begin{abstract}
There exists several modified gravity theories designed to reproduce the empirical Milgrom's formula (MOND). Here we derive analytical results in the context of the static weak-field limit of two of them (BIMOND, leading for a given set of parameters to QUMOND, and TeVeS). In this limit, these theories are constructed to give the same force field for spherical symmetry, but their predictions generally differ out of it. However, for certain realizations of these theories (characterized by specific choices for their free functions), the binding potential-energy of a system is increased, compared to its Newtonian counterpart, by a constant amount independent of the shape and size of the system. In that case, the virial theorem is exactly the same in these two theories, for the whole gravity regime and even outside of spherical symmetry, although the exact force fields are different. We explicitly show this for the force field generated by the two theories inside an elliptical shell. For more general free functions, the virial theorems are however not identical in these two theories. We finally explore the consequences of these analytical results for the two-body force.
\end{abstract}

\maketitle

\section{Introduction}
The current dominant paradigm is that galaxies are embedded in halos of cold dark matter. However, one observes that for gravitational accelerations below $a_0 \sim 10^{-10}$~ms$^{-2}$, the total gravitational attraction $g$ in galaxy disks approaches $(g_N a_0)^{1/2}$ where $g_N$ is the usual Newtonian gravitational field as calculated from the observed distribution of baryonic matter. The successes of this recipe in galaxies could be an emergent phenomenon, linked with the complex feedback between baryons and cold dark matter, but a more radical explanation of these successes is a modification of gravity on galaxy scales: this paradigm is known as modified Newtonian dynamics \citep[MOND, ][]{Mil83}. More precisely, within this paradigm, the Newtonian acceleration $\vec{g}_N$ produced by the visible matter is linked to the true gravitational acceleration $\vec{g}$ by means of an interpolating function $\mu(x)$:
\begin{equation}
\mu\left(g/a_{0}\right)\vec{g} = \vec{g}_{N},
\label{eq:1}
\end{equation}
where
$\mu(x) \sim x$ for $x \ll 1$ and
$\mu(x) \sim 1$ for $x \gg 1$ (and $g=|\vec{g}|$), or equivalently by means of an interpolating function $\nu(y)$:
\begin{equation}
\vec{g} = \nu\left(g_N/a_{0}\right)\vec{g}_{N},
\label{eq:2}
\end{equation}
where
$\nu(y) \sim y^{-1/2}$ for $y \ll 1$ and
$\nu(y) \sim 1$ for $y \gg 1$. However, these expressions cannot be exact outside of spherical symmetry, since they do not respect usual conservation laws. There exists various flavors of modified gravity theories reproducing this relation in spherical symmetry, but all making slightly different predictions outside of it.

For instance, in the Newtonian static weak-field limit of the generalized Einstein-Aether theories\cite{Zlosnik,Halle,Zhao2008}, the gravitational potential $\Phi$ obeys the following modified Poisson equation\cite{BM84}:
\begin{equation}
\nabla \cdot \left[ \mu(|{\mathbf\grad} \Phi|/a_0) {\mathbf\grad} \Phi \right]=4\pi G\rho.
\end{equation}
On the other hand, in Bekenstein's Tensor-Vector-Scalar (TeVeS) multifield theory\cite{Bek04}, the gravitational potential in the static weak-field limit can be expressed as:
\begin{equation}
\Phi=\Phi_N+\phi,
\end{equation}
where $\Phi_N$ is the Newtonian potential obeying the usual Poisson equation, and $\phi$ is a scalar field obeying an equation similar to Eq.~(3), but with a different $\mu$-function \citep[see, e.g.,][]{F07}, which can  simply be, at least far from the strongest gravity regime, $\mu(x)=x$. Finally, another possibility to recover Eqs.~(1) and (2) in the spherically symmetric static weak-field limit is the new BIMOND theory \cite{Mil09_bim,Clif} where we can have $\Phi=\Phi_N+\phi$, as in Eq.~(4), with $\phi$ obeying \citep[see also QUMOND,][]{Mil09_qu}:
\begin{equation}
\nabla^2 \phi=\nabla \cdot \left[ \nu(|{\mathbf\grad} \Phi_N|/a_0) {\mathbf\grad} \Phi_N \right], 
\end{equation}
where, e.g., $\nu(y) = y^{-1/2}$ far away from the strongest gravity regime.

Here, we intend to compare these various implementations of the MOND paradigm at the purely theoretical level, in the same philosophy as \cite{Dai,Matsuo}. A useful way to compare the theoretical implications of these various theoretical frameworks is to compare the virial theorem ensuing from them. Indeed, the virial theorem is a very useful tool, e.g. to compute the 2-body forces in these theories. The scalar form of the virial theorem has been computed for Eq.(3) in the deep-MOND limit by \cite{Mil94, Mil97} and in the deep-QUMOND case \cite{Mil09_qu}. Hereafter, we extend this study to the whole intermediate regime (not only deep-MOND) in the BIMOND and TeVeS frameworks, and show that for peculiar choices of the $\mu$ and $\nu$ functions, the scalar form of the virial theorem is identical in these two theories and independent of the size and shape of the system (Sect.~II), although the exact force fields are different. We explicitly show how the force fields differ in the case of a test particle sitting inside an elliptical shell of matter (Sect.~III). We also explore the consequences for the 2-body force (Sect.~IV).

\section{Virial Theorem}


A time-independent system in any gravitational theory satisfies  the tensor virial theorem:
\begin{equation}
2K_{jk} + W_{jk} = 0,
\end{equation}
where $K_{jk}=1/2 \int \rho \langle v_j v_k \rangle d^3r$ is the kinetic-energy tensor ($\rho$ being the density and $v_j$ being the j-th component of the velocity) and $W_{jk} = \int \rho r_j g_k d^3r$ is the Chandrasekhar potential-energy tensor (where $r_j$ the j-th component of the position and $g_k$ is the k-th component of the gravitational force). The form of this potential-energy tensor depends on the gravitational theory through $g_k$, and we explore herafter its form in the BIMOND and TeVeS frameworks. The trace of this tensor equation yields the scalar virial theorem.

\subsection{BIMOND}

In the non-relativistic quasi-static weak field limit of BIMOND, and for a given set of parameters, the gravitational potential $\Phi$ obeys \citep[see Eq.54 of][]{Mil09_bim} 
$\nabla^2 \Phi=4 \pi G \rho + \nabla \cdot \left[ \nu(|{\mathbf\grad} \Phi_N|/a_0) {\mathbf\grad} \Phi_N \right]$.
This means that the potential $\Phi$ can be divided into a Newtonian and a quasi-linear MOND (QUMOND) part as in Eqs.~(4) and (5), and that in the potential-energy tensor, each $g_k$ can be decomposed into a Newtonian part and a QUMOND part $g_k=g_{N_k} + g_{Q_k}$, so that the scalar virial theorem ensuing from the trace of Eq.~(6) writes, for the whole gravity regime, $2K+W_N+W_Q=0$.

In order to recover the asymptotic form $\nu \sim \sqrt{a_0/g_N}$ for $g_N \ll a_0$ in Eq.~(2) for the total force (Newton+QUMOND) in the spherical BIMOND case, the (different) $\nu$-function of the modified Poisson equation (Eq.~5) must also have the asymptotic form  $\nu(y) = y^{-1/2}$ (where $y=g_N/a_0$) for $r \rightarrow \infty$. Then, as shown by Milgrom \cite{Mil09_qu}, we have
\bey
& &-W_Q = {2 \over 3} (GM^3a_0)^{1/2} + \int Z \beta d^3r, 
\\ & & Z(\nu^{-1}) \equiv {\nu \vec{g}_N(r) \cdot \vec{g}_N(r)  \over 8\pi G},~ \beta \equiv Z^{-1} \int_0^Z  dZ \left[ {d\ln (Z \nu^3) \over d\ln Z} \right] \nonumber
\eey
where $Z$ and $\beta$ are functions of $\nu^{-1}$, $Z$ is an energy density, and $\beta$ is a dimensionless number depending on nothing except on the shape of the $\nu$-function.



\subsection{TeVeS}

Generally speaking, for  a curl-free vector $\vec{s} = (s_1,s_2,s_3)$, a function $F=F(s^2)$ (where $s\equiv|\vec{s}|$), and an arbitrary vector $\vec{u}=(u_1,u_2,u_3)$, we have the following identity:
\bey\label{S3d}
 (s_j u_j) S  &=&   \partial_i \left[ F' s_i s_j u_j -   {F u_i \over 2} \right] \\ 
& &+ \left[  {(\partial_j u_j) F \over 2} -  F' s_j s_i  {\partial_i u_j + \partial_j u_i \over 2} \right].
\nonumber
\eey
where $S \equiv \partial_i ( F' s_i)$, and where we adopted the notation of implicitly summing over common indexes $i$ or $j=1,2,3$, and we used the fact that 
$\partial_i s_j = \partial_j s_i$ because of curl-freeness.
Integrating over the entire 3D volume, the first term on the rhs becomes a surface integral through the divergence theorem.  

Now, for a TeVeS-like multifield theory, the static weak-field gravitational potential can be separated into a Newtonian part $\Phi_N$ and a scalar field MONDian part $\phi$, as in Eq.~(4). This makes it similar to BIMOND and different from the classical MONDian potential that fully obeys an Equation like Eq.~(2). Again, in the potential-energy tensor, each $g_k$ can be decomposed, so that the virial theorem of Eq.~(6) becomes
$2K_{jk} + W_{N_{jk}} + W_{s_{jk}} = 0$,
where $W_{s_{jk}}= -\int \rho r_j \, \partial_k \phi \, d^3r$. We can now identify in the above identity (Eq.~8), $\vec{s}=-\grad \phi/a_0, \, F'= \mu, \, S = -4 \pi G \rho/a_0$, where $\phi$ is the scalar field of Eq.~(4), $\rho$ is the matter volume density, whose integration over the whole volume is the mass $M$, and $\mu$ is the interpolating function of Eq.~(3), which the scalar field $\phi$ obeys.

Let now $u$ be a constant vector $\vec{u}=(1,0,0)$, then integrating Eq.~(\ref{S3d}) gives the total scalar force's 1-component over the system $(a_0^2/4 \pi G) \int \left[ F' {\vec{s}\cdot d\vec{A}} s_1  -   (F/2) dA_1 \right]$.
On the other hand, let $u$ be a divergence free field $\vec{u}=(0,-r_3,r_2)$, then Eq.~(\ref{S3d}) gives the total scalar torque's 1-component over the system (or the rate of change of the angular momentum along the 1-axis due to the scalar field), $a_0^2/4\pi G \int  \left[ F' {\vec{s}\cdot d\vec{A}} (r_2 s_3-r_3 s_2) \right]$. Also, adopting $\vec{u}=\vec{s}$ and $\vec{u}=\mu \vec{s}$ we obtain the following ``theorems" for the scalar field
\bey
\overline{s^2} &=& {1 \over M} \int dr^3 \rho_{\rm eff} F \\
\overline{2 s^2 F' -F} &=& {1 \over M} \int dr^3 \rho_{\rm eff} (F's)^2,
\nonumber
\eey
where $\rho_{\rm eff}$ is the effective density of matter that would have sourced the scalar field in Newtonian gravity.
These formulae, valid in any geometry and for any $\mu$ (and $F$) functions, are useful for checking the self-consistency of a numerically computed $\rho_{\rm eff}$ and the corresponding MONDian field $s$.

Finally, if $\vec{u}$ is a vector $\vec{u}=(r_1,0,0)$, then Eq.~(\ref{S3d}) gives the (1,1)-component of the Chandrasekhar potential energy-tensor $W_{s_{11}}$ of the scalar field in the system 
\bey
& &-{4\pi G \over a_0^2} W_{s_{11}} =  \\
& & \int \left[ (\vec{s} \cdot d \vec{A}) r_1 s_1 F'  -   {F r_1 d A_1 \over 2}  \right] 
+ \int d^3 r \left[  {F \over 2}  -  F' s_1 s_1  \right].
\nonumber
\eey
The sum $W_{s_{11}} + W_{s_{22}} +W_{s_{33}}$ then leads to the scalar virial theorem in the quasi-static weak-field limit of TeVeS, $2K + W_{N}  + W_{s}=0$, where, similarly to BIMOND, $W_{s}=W_{s_\infty}+W_{s_{\mu}}$, $W_{s_\infty}$ corresponding to the sum of the first terms of the rhs of Eq.~(10) (taken as a surface integral at infinity), and $W_{s_{\mu}}$ corresponding to the sum of the second terms of the rhs of Eq.~(10):
\beq
-W_{s_{\mu}} = \int  Z \beta d^3r , \, Z(\mu) \equiv {\mu \vec{g}_s(r) \cdot \vec{g}_s (r)  \over 8\pi G}.
\eeq

In order to recover the asymptotic form $\mu \sim g/a_0$ for $g \ll a_0$ in Eq.~(1) for the spherical case in TeVeS, the (different) $\mu$-function of the modified Poisson Eq.~(3) obeyed by the scalar field must always have the asymptotic form $F'=\mu(s) = s$ for $s \ll 1$ (and $F(z) = (2/3) z^{3/2}$ where $z=s^2$) \citep[see Eq.27 in][]{F07}. The surface integral $W_{s_\infty}$ thus becomes a constant because of this asymptotic behavior at infinity \citep[see][]{Mil94}
$-W_{s_{\infty}} 
= {2 \over 3} (GM^3 a_0)^{1/2}$.
Quite remarkably, this result combined with Eq.~(7) and Eq.~(10) directly shows that, if $\nu=\mu^{-1}$, and if $g_N=\mu g_s$ (i.e. in spherical symmetry), the virial theorems are precisely identical in the two theories TeVeS and BIMOND. This was to be expected since they were constructed to give the same result for the total force field in spherical symmetry. Outside of spherical symmetry, $g_N \neq \mu g_s$ and the virial theorems are thus different.

Let us note that for such an asymptotic behaviour of the $\mu$-function (and $F$-function), the total scalar force's 1-component (see the paragraph before Eq.~9) is zero, and the momentum 
along the 1-axis (and along the other axes) of the system is conserved.  Similarly, the cancellation of the total scalar torque for this asymptotic behavior of $F$ implies that the total angular momentum 
along each axis is also conserved.

\subsection{A special case}

Besides this asymptotic behaviour of the $\mu$-function for $s \ll 1$, there are many different forms of the $\mu$-function for the scalar field in the intermediate to strong gravity regime, like e.g. $\mu \rightarrow \infty$ for $s \rightarrow 1$ \citep[see][]{F07}. In BIMOND, different forms of $\nu=\mu^{-1}$ are thus also possible. However, a possibility is to choose $\mu(s) = s$ everywhere \cite{Bek04}, which, at least for the intermediate gravity regime of galaxies and galaxy clusters, can fit observations \cite{F07} (with an additional dark matter component in galaxy clusters \cite{AFD}). In BIMOND, this corresponds to $\nu(y)=y^{-1/2}$.

In this special case, the volume integral of Eq.~(7) goes to zero since $g_N \sim \nu^{-2}$ and $Z \sim \nu^{-3}$, meaning that $\beta=0$. In TeVeS, the volume integral $W_{s_{\mu}}$ (see Eq.~10 and after) also cancels since $Z \sim \mu^3$, and $\beta$ is also zero since $\mu=\nu^{-1}$. This remarkable result, corresponding to the deep-MOND result of Milgrom \cite{Mil94,Mil09_qu}, means that, for this choice of the $\mu$ and $\nu$ functions, independently of the size and shape of the system, the potential-energy of any system becomes more negative, compared to its Newtonian counterpart, by a constant amount $-(2/3)\sqrt{GM^3a_0}$, both in BIMOND and TeVeS. However, in this special case, while the total potential energies and virial theorems have the remarkable property of being identical in the two theories, the exact force fields are still different. We explicitly show this in the next section in the case of elliptical shells.

\section{Elliptical shells}

The above specific choice of $\mu$ and $\nu$ functions, for which the total potential energy of any system is identical in TeVeS and BIMOND, corresponds to a force field that can be separated into, on one hand, a Newtonian part, and, on the other hand, a ``deep-MOND'' or a ``deep-QUMOND'' part, respectively. In order to compare the force fields generated by these two implementations of the theories, we can thus concentrate on the differences between the predictions of MOND and QUMOND in the deep-MOND and deep-QUMOND regimes. 

An interesting way of comparing the deep-MOND and deep-QUMOND predictions is to consider, for a system in a low gravity regime, the effective density of matter $\rho_{\rm eff}$ that would have sourced the force field in Newtonian gravity: $
\rho_{\rm eff}^{\rm MOND} \equiv  (1 / 4 \pi G) \nabla^2 \Phi,  \;
\rho_{\rm eff}^{\rm QUMOND} \equiv  (1 / 4 \pi G) \nabla.(\nu \nabla \Phi_N )$. In TeVeS and BIMOND, this additional gravity can be attributed to what observers would call ``dark matter'', and the above effective matter density is called ``phantom dark matter" density \cite{Milph,Wu}.

Let us now concentrate on the theoretical case of a system with all its mass in an elliptical shell. The Newtonian prediction is that the gravity is uniformly zero in the void inside the shell, hence $\nu \nabla \Phi_N$ is constant (zero), and QUMOND predicts
$\rho_{\rm eff}^{\rm QUMOND} =0$

In the case of classical MOND, however, this is not the case.  An example is shown in Fig.~1.  We consider a density model with the baryonic profile 
$\rho_b=0$ for $a<1$ and $\rho_b=a^{-4}(1-a^{-1})^2$ for $a \geq 1$, 
where $a^2 \equiv R^2+ {z^2 \over q^2}$ and $q=0.7$. We then use a modified Poisson solver to solve Eq.~3 for $\mu(x)=x$. We then compute the effective matter volume density contours in the $Rz$-plane (shaded areas on Fig.~1). It is clearly non-zero inside the shell, meaning that, even though the total potential energy of the system is the same in deep-MOND and deep-QUMOND, the force fields are different. In fact, the predictions of the two theories can be decomposed into their prediction in spherical symmetry (that are the same by construction) and a curl-field: it is this curl-field which differs in MOND and QUMOND.

In general, the QUMOND effective density is more intuitively related to the Newtonian gravity than in classical MOND. For instance, in QUMOND, whenever the Newtonian gravity is locally a central force field, the effective density in that region is locally spherical (this is however not true for a system plunged into a uniform external force field, except if the internal Newtonian gravity is also uniform in the region of interest).

\begin{figure}
\centerline{\includegraphics[angle=0,width=9cm]{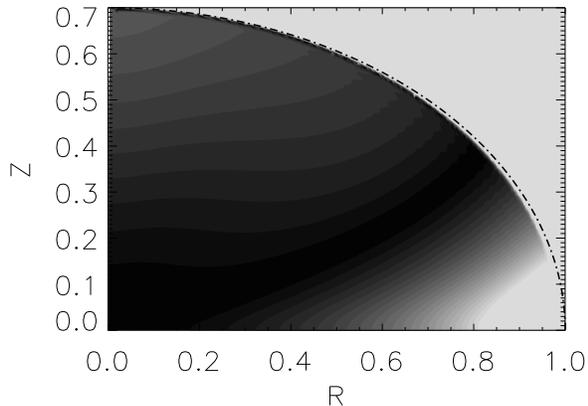}}
\caption{Effective density predicted by (deep-)MOND inside an elliptical shell (see Sect.~III; the baryonic density is zero inside the dot-dashed line). The contours are for $|r^2\rho_{\rm eff}|=0$ (black) to $|r^2\rho_{\rm eff}|=10^{-3}$ (white) in steps of $10^{-4}$. The dark band in the middle indicates zero density, the upper grey areas have negative densities, and the lower ones positive. In QUMOND, this effective density is zero inside the shell.}
\end{figure}

\section{Conclusion and discussion}

We conclude that, in general, the TeVeS and BIMOND force fields and virial theorems are different, except in spherical symmetry (with $\nu=\mu^{-1}$). However, for a specific choice of the $\mu$ and $\nu=\mu^{-1}$ functions (observationally only valid in the intermediate gravity regime), the virial theorems are identical even outside of spherical symmetry. What is more, the potential energy of the system can then be expressed analytically, for the whole gravity regime, as a sum of the Newtonian potential energy and of a constant term proportional to the 3/2 power of the total mass of the system. However, while the virial theorems are then identical, the exact force fields are still different (i.e. the gravities have different curl-fields), as we showed in the case of elliptical shells.

In this specific case, we can also get an analytic expression for the 2-body force under the approximation that the two bodies are very far apart compared to their internal sizes \citep[see also][]{Mil94}. Since $2K+W_N-(2/3)\sqrt{GM^3a_0}=0$, since the kinetic energy can be separated into the orbital energy $K_{\rm orb}=M_1M_2 v^2_{\rm rel}/(2M)$ and the internal energy of the bodies $K_{\rm int}=-\Sigma (1/2)W_{N_i}+\Sigma (1/3) \sqrt{GM_i^3a_0}$, and since the Newtonian potential energy can be separated into the interaction term (the mutual potential energy) and the internal Newtonian potential energies of the bodies $\Sigma W_{N_i}$, we get:
\beq
\frac{M_1M_2 v^2_{\rm rel}}{M} = \frac{GM_1M_2}{r_{12}}+\frac{2}{3} \left[ (GM^3a_0)^{1/2} - \sum_i (GM_i^3a_0)^{1/2} \right].
\eeq
We can then assume an approximately circular velocity such that the 2-body force can be written
\bey
& & \vec{F}_{12} =  M_1\vec{a_1}=\frac{M_1M_2v^2_{\rm rel}\vec{r}_{12}}{Mr^2_{12}} = \frac{GM_1M_2\vec{r}_{12}}{r^3_{12}} \\
& & + \frac{2}{3}\left[ 1 - \sum_{i=1}^{i=2} \left(\frac{M_i}{M}\right)^{3/2} \right]\frac{(GM^3a_0)^{1/2}\vec{r}_{12}}{r^2_{12}}. \nonumber
\eey
For this specific choice of free function, the TeVeS and BIMOND 2-body forces are thus the same, provided the two bodies are very small compared to their mutual distance, but if they are not, the force will be different since we showed that the force fields are different in general. As a final remark, let us stress that this 2-body force depends on the mass ratio of the binary, meaning that the orbital period of an equal-mass binary would differ from that of a binary of extreme ratio but with the same total mass and separations, which could cause stars to segregate according to their masses.  

\begin{acknowledgments}
The authors thank X. Wu for her help. HSZ acknowleges hospitality at the {\it Observatoire de Strasbourg}.
\end{acknowledgments}

\end{document}